\documentclass[
aps,showpacs
 ,twocolumn
]{revtex4}
\usepackage{tikz}
\usepackage{adjustbox}
\usetikzlibrary{arrows, hobby}
\usepackage{wrapfig}
\usepackage{graphicx}
\usepackage{amsmath}
\usepackage{amssymb}
\usepackage{hyperref}
\usepackage{ mathrsfs }
\usepackage{indentfirst}
\usepackage{braket}
\usepackage{color}
\usepackage{empheq}
\usepackage{lipsum,adjustbox}
\usepackage[outdir=./]{epstopdf}
\usetikzlibrary{calc,positioning}

\tikzstyle{beamsplitter}=[fill=blue, fill opacity=0.2]

\begin{document}

\title{Quantum limited measurement of space-time curvature with scaling beyond the conventional Heisenberg limit}

\author{ S. P. Kish and T. C. Ralph}\affiliation{Centre for Quantum Computation and Communication Technology, \\
	School of Mathematics and Physics, University
	of Queensland, Brisbane, Queensland 4072, Australia}

\date{\today}

\begin{abstract}
	{}
	We study the problem of estimating the phase shift due to the general relativistic time dilation in the interference of photons using a  non-linear Mach-Zender interferometer setup.   
	By introducing two non-linear Kerr materials, one in the bottom and one in the top arm, we can measure the non-linear phase $\phi_{NL}$ produced by the space-time curvature and achieve a scaling of the standard deviation with photon number ($N$) of $1/N^{\beta}$ where $\beta > 1$, which exceeds the conventional Heisenberg limit of a linear interferometer ($1/N$). 
	The non-linear phase shift is an effect that is amplified by the intensity of the probe field. In a regime of high photon number, this effect can dominate over the linear phase shift.  
	
\end{abstract}


\maketitle


\vspace{10 mm}

Metrology is a key driver of technology. Ultimately, however, the ability to estimate parameters of physical systems is restricted by quantum mechanics. Quantum metrology studies how the fundamental bounds on the resolution of such estimates depend on resources such as energy \cite{GIO11}. It is hoped that such studies will lead to new techniques allowing the development of measurement devices of unprecedented precision.

For example, the use of a laser probe to measure a phase-shift, $\theta$, is fundamentally limited by the quantum noise of the probe coherent state. The standard deviation of the estimate, $\langle \Delta \theta \rangle$, scales with the average photon number of the probe states, $N$, as $\langle \Delta \theta \rangle \propto1/\sqrt{N}$. This is known as the standard quantum limit. Very high laser powers are used in gravitational wave interferometers to exploit this scaling \cite{ABB16}. It is well known that a squeezed state probe can do better, leading ideally to a $\langle \Delta \theta \rangle \propto 1/N$ scaling known as the Heisenberg limit \cite{CAV81}. Achieving the Heisenberg limit under practical conditions is extremely demanding. 

Recently it has been observed that if there is a strong non-linear coupling to the probe then energy scalings better than the conventional Heisenberg limit can be achieved \cite{BOI07, BOI}. These claims have generated some controversy \cite{HAL12,zwierz}. Never-the-less a spin-based experimental system has been demonstrated \cite{NAP11}. In the optical domain an example is that of probe transmission through a Kerr medium where it has been shown that estimation of the non-linear parameter, $\chi$, can be achieved with a $\langle \Delta \chi \rangle \propto 1/N^{3/2}$ scaling \cite{JOO12}. Whilst this is intriguing, there have been few proposed applications for such an effect \cite{LUIS}.  Normally we would be interested in estimating some external parameter -- not the strength of the measurement system non-linearity itself. 

%

In this paper we note that, due to time dilation, the effective non-linearity of a fixed length of a non-linear medium is a function of the local gravitational field. This is in addition to the linear phase that is also a function of the proper time. We use this effect to construct an interferometric arrangement that allows one to estimate the space-time curvature of the field  with a scaling beyond the conventional Heisenberg energy limit of a linear interferometer \cite{zych}. Current techniques for measuring gravity such as atom interferometry \cite{peters} are limited to the standard quantum limit (SQL). Squeezing and entanglement could enhance the performance of atom interferometers  \cite{savas, szigeti, esteve, gross} but only up to the Heisenberg limit.

Consider light propagating through a Kerr non-linearity in a gravitational field described by the 
Schwarzschild metric. We assume that the metric is approximately constant over the length of the medium. The Kerr non-linearity constant $\chi$ is coupled to the proper time $\tau$ it takes to interact with the medium, as measured locally \cite{qft}. Thus the effective non-linearity becomes $\chi'=\chi \tau$. This essentially means that the effective non-linearity depends on the curvature of space-time. For a non-linearity of length $L$, the proper time as measured by an observer at radius $r=r_0$, relative to some reference observer situated at a different radius, is $\tau \approx (1-\frac{K r_s}{2 r_0}) \frac{L}{c}$ where $r_s=\frac{2 G M}{c^2}$ is the Schwarzschild radius and $K$ is a constant that depends on the position of the reference observer. We can see that the non-linear coupling is approximately proportional to the Schwarzschild radius.  The stronger the curvature $r_s$, the stronger the space-time coupling to the non-linearity. In principle we can estimate the spacetime curvature using this dependence.   
%

We model the transmission of a coherent state probe with amplitude $\alpha$ through the medium as the unitary evolution $\ket{\alpha_{NL}(\tau)} = \hat{U} \ket{\alpha}$ where $\hat{U}=
e^{i \chi \tau \hat{n} (\hat{n}+1)+i\hat{n} k c \tau}$ with $\hat n$ the number operator, and $k$ the wave number of the optical mode \cite{milburn}. Hence we find:
\begin{equation}
\ket{\alpha_{NL}(\tau)}=e^{-|\alpha|^2/2} \sum_{n=0}^{\infty} \frac{(\alpha e^{i \chi \tau (n+1)+i k \phi(\tau)})^n}{\sqrt{n!}}  \ket{n}
\label{coh}
\end{equation}
We want to determine the ultimate quantum bound for estimating $r_s$ using non-linear couplings. The bound for the variance of an unbiased estimator $\hat{\tau}$ is determined by the Cramer-Rao inequality \cite{cram}. In quantum information theory, for $M$ number of independent measurements, the inequality is $\braket{\Delta \hat{\tau}^2} \ge \frac{1}{M \mathcal{H}(\tau)}$. Where $\mathcal{H}(\tau)$ is the Quantum Fisher Information which represents the most information obtainable by a parameter for an optimal quantum measurement \cite{mras}. This type of analysis determines the local precision \cite{HAL12} i.e. it assumes we start with a good initial estimate of $r_s$, which we seek to refine. 

We determine the Quantum Fisher Information via \cite{introqfi, qfidef, BRA94, safranek, strobel}:
%
%
\begin{equation}
\mathcal{H}(\tau)=\lim_{d\tau \to 0} \frac{8(1-\sqrt{\mathcal{F}(\rho(\tau),\rho(\tau+d\tau))})}{d\tau^2}
\label{qfib}
\end{equation}
where $\mathcal{F}(\rho,\sigma)=(Tr(\sqrt{\sqrt{\rho} \sigma \sqrt{\rho}}))^2$ is the quantum fidelity between two density matrices $\rho$ and $\sigma$.
We want to determine the QFI for the probe coherent state undergoing the non-linear evolution (Eq. \ref{coh}). We disregard orders higher than 2 in $d \tau$ as $d \tau \rightarrow 0$ and $N_a$ is finitely large. Therefore we find the modified fidelity is (see Appendix A for calculation of overlap) 
%
%

%
\begin{equation}
\begin{split}
\mathcal{F}&= |\braket{\alpha_{NL}(\tau+d \tau)| \alpha_{NL}(\tau)}|^2\\
&=1-d \tau^2 N_a (2(2+5 N_a+2 N_a^2) \chi^2+4 (1 +N_a) \chi \omega+\omega^2)
\end{split}
\end{equation}
and hence:
\begin{equation}
\begin{split}
\mathcal{H}(\tau)
&=4 N_a ((\omega+2(N_a+1)\chi)^2+2 N_a \chi^2)
\end{split}
\label{qfi}
\end{equation}
Where $\omega=k c$ is the frequency and $N_a=|\alpha|^2$ is the photon number of the single mode. By noting that $\mathcal{H}(r_s)=(\frac{d \tau}{d r_s})^2 \mathcal{H}(\tau)$ and $\frac{d\tau}{d r_s}=\frac{-K L}{2 c r_0}$, we find the relative error of the space-time parameter $r_s$ is given by:
\begin{equation}
\frac{\braket{\Delta r_s}_{opt}}{r_s} \ge \frac{c r_0}{ K L r_s \sqrt{ N_a ((\omega+2(N_a+1)\chi)^2+2 N_a \chi^2)} }
\label{CRbound}
\end{equation}
For $N_a$ large we see the scaling beyond the conventional Heisenberg limit of the relative error.

We can generalize this result for the case of higher non-linearities where the light that propagates through a non-linear media experiences self-interaction described by the general Hamiltonian: $\hat{H}=\chi (a^\dagger a)^q$. Where $q \ge 2$ and $\chi$ is a coupling constant. 
%
%
For large $N_a$ the relative error of the parameter $r_s$ is given by (see Appendix B):
\begin{equation}
\frac{\braket{\Delta r_s}_{opt}}{r_s} \ge \frac{c r_0}{ K L r_s \sqrt{ N_a (q \chi N_a^{q-1}+\omega)^2} }
\end{equation}
Clearly, the standard deviation of the space-time parameter scales as $\braket{\Delta r_s}_{opt} \propto \frac{1}{ q \chi N_a^{\frac{2q-1}{2}}}$. Since the time dilation is coupled to the non-linearity, when $q \chi N_a^{q-1} >> \omega$, it is advantageous to measure the non-linear phase rather than the linear phase. 


\tikzstyle{block} = [draw, fill=blue!20, rectangle, 
minimum height=3em, minimum width=6em]
\tikzstyle{sum} = [draw, fill=blue!20, circle, node distance=1cm]
\tikzstyle{input} = [coordinate]
\tikzstyle{output} = [coordinate]
\tikzstyle{pinstyle} = [pin edge={to-,thin,black}]
\begin{figure}
	\begin{center}
		\includegraphics[width=\linewidth]{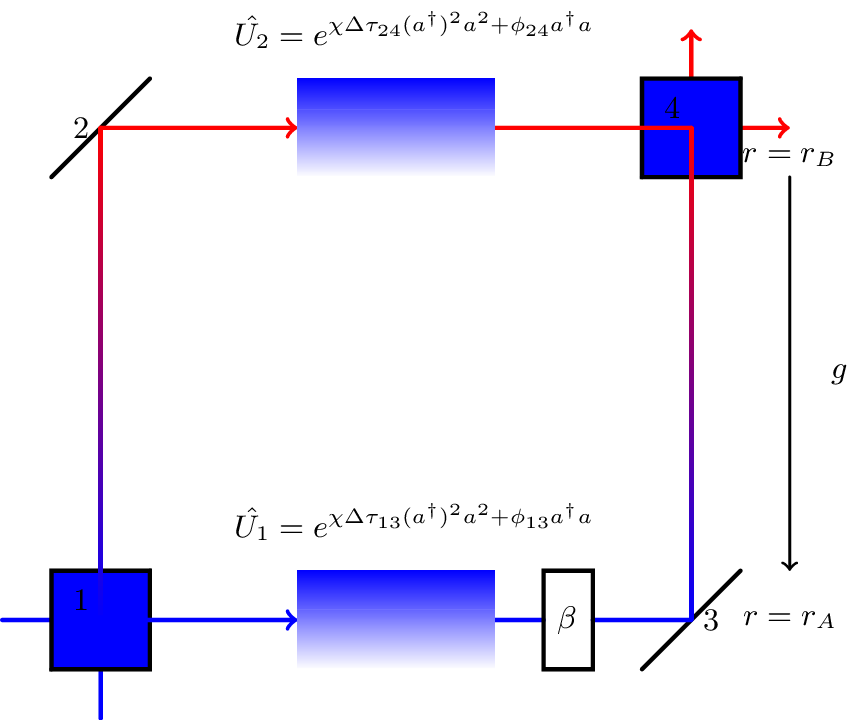}
	\end{center}
	\caption{Non-linear interferometer of arm length $L$ in a gravitational field. Coherent light passes through a $50/50$ beamsplitter at 1. The phase from 1 to 3 at $r=r_A$ is set to $\phi_{13}=0$ and the time interval that light traverses is $\tau_{13}=\frac{L}{c}$. The effect of the gravitational redshift cancels out and no phase shift is imposed as light traverses vertically. In the top and bottom arms, we have a non-linear medium with $\chi$ coupling. A phase difference due to slower interaction time with the non-linearity in the bottom arm is detected after recombining at the second beamsplitter. The time intervals $\Delta \tau_{13}$ and $\Delta \tau_{24}$ contain the Schwarzschild radius $r_s$. $\beta$ represents an adjustable linear phase shift.}
	\label{int}
\end{figure}

{\em A non-linear interferometer} - We now propose a device for realising the enhanced sensitivity suggested by Eq. \ref{CRbound}. We consider the Mach-Zender interferometer shown diagrammatically in Fig \ref{int}. We describe the gravitational field via the Schwarzschild metric with line element $ds^2=g_{\mu \nu}dx^\mu dx^\nu=-f(r)dt^2+\frac{1}{f(r)} dr^2 + r^2 d\phi^2$ where $f(r)=1-\frac{r_s}{r}$. An observer at a fixed radius $r=r_0$ will measure the proper time $\tau=\int ds= \sqrt{f(r_0)} t$ where $t$ is the proper time measured by an observer at infinite distance $r=\infty$. Without loss of generality we have assumed we are in the equatorial plane with $d\phi$ the usual angular coordinate. Let us first consider evolution of a probe state through the interferometer in Fig \ref{int} without the Kerr non-linearities.  

%
The output modes can be written in terms of the input modes as \cite{zych}:
\begin{equation}
\begin{split}
b_k&=\frac{1}{2}(a_k (e^{-ik(\phi_{12}+\phi_{24})}-e^{-ik(\phi_{13}+\phi_{34})})\\
&+v_k(e^{-ik(\phi_{12}+\phi_{24})}+e^{-ik(\phi_{13}+\phi_{34})}))
\end{split}
\end{equation}
where $a_k$ is prepared in the coherent state and $v_k$ in the vacuum state.
The phase shifts in the vertical arms are equal and so cancel out. Therefore we can set $\phi_{12}=\phi_{34}=0$ without loss of generality. 
%
%
%
%
%
In the bottom horizontal arm, we can choose the time interval so that the phase $\phi_{13}=0$ and thus $\Delta x_{r_A,13}=c \Delta \tau_{r_A,13}$. We are assuming that $\Delta x_{r_A,13}$ is sufficiently small that we can disregard the curvature of space-time in the horizontal direction. 
The unknown is $\phi_{24}=\Delta x_{r_B,24}-\frac{c}{n'} \Delta \tau_{r_B,24}$, where $n'$ is the first order refractive index of the material. In the Schwarzschild metric, the proper time interval at $r=r_A$ is $c \Delta \tau_{r_A}= c \sqrt{1-\frac{r_s}{r_A}} \Delta t= \Delta x_{r_A}$,
where $\Delta t$ is the time interval as seen by a far-away observer, and $r_s=\frac{2GM}{c^2}$ is the Schwarzschild radius. We also know that at $r=r_B$ the proper time is 
\begin{equation}
\frac{c}{n'} \Delta \tau_{r_B}= \frac{c}{n'} \sqrt{1-\frac{r_s}{r_B}} \Delta t=\frac{\sqrt{1-\frac{r_s}{r_B}}}{n' \sqrt{1-\frac{r_s}{r_A}}} \Delta x_{r_A}.
\label{time}
\end{equation}
Since the length of the top arm is the same as the bottom arm we set: $\Delta x_{13}=\Delta x_{24}=L$, and to simplify the nomenclature we redefine $\tau_{13}=\tau_1$ and $\tau_{24}=\tau_2$: $\tau_{2}=\frac{\sqrt{1-\frac{r_s}{r_B}}}{\sqrt{1-\frac{r_s}{r_A}}} \tau_1 \approx (1- \frac{r_s h}{2 r_A r_B}) \frac{L}{c}=(1-\delta) \frac{L}{c}$,
where we have defined $\delta= \frac{r_s h}{2 r_A r_B} $. This approximation assumes $r_{A,B}>>r_s$. 
The {\em linear} phase simplifies to:
\begin{equation}
\phi_{24}= L-\frac{c}{n'} \tau_2=(1-\frac{\sqrt{1-\frac{r_s}{r_B}}}{n' \sqrt{1-\frac{r_s}{r_A}}})L\approx (1-\frac{1}{n'}+\frac{r_s h}{2 r_A r_B n'}) L
\label{phase}
\end{equation}

Now we place two non-linear Kerr media in the top and bottom arms, we expect a phase shift due to the same time dilation, but the Kerr non-linear medium induces an additional intensity dependent phase shift. 
The Heisenberg evolution of the annihilation operator for the Kerr non-linear effect is $a_k (\tau)=e^{i\chi \tau a_k^{\dagger} a_k} a_k$ \cite{dodonov}. Thus the output mode of the Mach-Zender non-linear interferometer is given by:
\begin{equation}
\begin{split}
b_k=&\frac{1}{2}((e^{-ik \phi_{2}+i \chi \tau_{2} a_{k}^{\dagger} a_{k}}-e^{-ik\phi_{1}+i \chi \tau_{1} a_{k}^{\dagger} a_{k}+i \beta})a_k \\
&+(e^{-ik\phi_{2}+i  \chi \tau_{2}a_{k}^{\dagger} a_{k}}+e^{-ik\phi_{1}+i \chi \tau_{1} a_{k}^{\dagger} a_{k}+i\beta})v_k)
\end{split}
\label{output}
\end{equation}
%
%
%
We know from Eq. \ref{time} and \ref{phase} the measured proper time $\tau_{2}$ and the phase $\phi_{2}$.
%
The time intervals $\tau_{1}$ and $\tau_{2}$ contain the Schwarzschild radius $r_s$. We also include an additional adjustable linear phase shift, $\beta$.





{\em Estimating the space-time curvature}.- To achieve the optimal error bound, we need to make an appropriate measurement at the interferometer output. We assume the coherent amplitude of the probe is large enough to treat as a classical coherent amplitude with added vacuum fluctuations which are only retained to first order. Hence writing $a = \alpha + \delta a$, this allows us to approximate the Kerr evolution in the following way: $e^{-i a^\dagger a \chi \tau} a \approx e^{-i |\alpha|^2 \chi \tau-i \chi \tau (\alpha^* \delta a+\alpha \delta a^\dagger)} (\alpha+\delta a) \approx e^{-i |\alpha|^2 \chi \tau} (1-i \chi \tau (\alpha^* \delta a+\alpha \delta a^\dagger))(\alpha+\delta a) =e^{-i |\alpha|^2 \chi \tau} (1-i \chi \tau (\alpha^* \delta a+\alpha \delta a^\dagger))\alpha +e^{-i |\alpha|^2 \chi \tau} \delta a$.
 This approximation is justified provided that $\tau \chi \alpha=\tau \chi \sqrt{N} <<1$. Unlike Ref. \cite{LUIS}, this is a looser restriction on the parameters $\tau$, $\chi$, and $N$. By remaining in the linearized Gaussian regime, it is a good approximation to work with single mode pulses \cite{shapiro, kit}. Thus, we continue our analysis in single modes. By applying this approximation to the interferometer mode at the output given by Eq. \ref{output}, we can write the approximate output quadrature amplitude at angle $\theta$ as:
\begin{equation}
\begin{split}
X_b&=b(\tau) e^{i\theta}+b^\dagger(\tau)e^{-i\theta}\\
&=|\alpha| \cos{(\theta+\zeta_2}) - |\alpha| \cos{(\theta+\zeta_1+\beta)} \\
&-\chi |\alpha|^2 (\tau_2 \sin{(\theta+\zeta_2)}-\tau_1 \sin{(\theta+\zeta_1+\beta)}) X  \\
&+\frac{1}{2}(X_{\theta+\zeta_2}-X_{\theta+\zeta_1+\beta})\\
& +\frac{1}{2}(X_{v(\theta+\zeta_2)}+X_{v (\theta+\zeta_1+\beta)})
\label{xb}
\end{split}
\end{equation}
where, to simplify the notation, we define $\zeta_1=k \phi_1-\tau_1 \chi |\alpha|^2$ and $\zeta_2=k \phi_2-\tau_2 \chi |\alpha|^2$ where $\tau_2 \approx (1-\delta) \tau_1$. We find $\braket{X_b}=|\alpha| (\cos{(\theta+\zeta_2)}-\cos{(\theta+\zeta_1+\beta)})$. Therefore, the dark port occurs at $\beta_{dark}=\zeta_2-\zeta_1$. Noting that $\frac{d \tau_2}{d r_s}=-\frac{\delta}{r_s}\frac{L}{c}$ and $\frac{d \tau_1}{d r_s}=0$ we find the derivative w.r.t. $r_s$ of the quadrature is $\frac{d \braket{X_b}}{d r_s}=-|\alpha| (\frac{k c}{n'}+|\alpha|^2 \chi)(\frac{d \tau_2}{d r_s}\sin{(\theta+\zeta_2)}-\frac{d \tau_1}{d r_s} \sin{(\theta+\zeta_1+\beta)}) =|\alpha| (\frac{k c}{n'}+|\alpha|^2 \chi) \frac{\delta L}{r_s c}\sin(\theta+\zeta_2))$.
%
%
The quadrature variance is given by
$\braket{\Delta X_b^2}=\chi^2 |\alpha|^4 (\tau_2 \sin{(\theta+\zeta_2)}-\tau_1 \sin{(\theta+\zeta_1+\beta)})^2-\chi |\alpha|^2 (\tau_2 \sin{(\theta+\zeta_2)}-\tau_1 \sin{(\theta+\zeta_1+\beta)}) \times (\cos{(\theta+\zeta_2)}-\cos{(\theta+\zeta_1+\beta)})+1$.

%
The effect of the non-linearity creates undesirable noise from anti-squeezing in the axis of rotation. However, we can optimize for our choice of $\beta$ to force the variance to be shot noise. More generally the solution is $\frac{\sin{(\theta+\zeta_2)}}{\sin{(\theta+\zeta_1+\beta)}}=\frac{\tau_1}{\tau_2}$ implying that we require $\beta=-\theta-\zeta_1+\arcsin{(\frac{\tau_2}{\tau_1}\sin{(\theta+\zeta_2)})}$. Furthermore, the derivative of the quadrature is  $|\alpha| (\frac{k c}{n'}+|\alpha|^2 \chi) \frac{\delta L}{r_s c} \sin(\theta+\zeta_1+\beta)(1+\frac{\tau_1}{\tau_2})=|\alpha| (\frac{k c}{n'}+|\alpha|^2 \chi) \frac{\delta L}{r_s c} (\frac{\tau_2}{\tau_1}\sin{(\theta+\zeta_2)})(1+\frac{\tau_1}{\tau_2})$. 
The optimal measurement angle is $\theta=\frac{\pi}{2}-\zeta_2$, and $\beta=\zeta_2-\zeta_1-\frac{\pi}{2}+\arcsin{(\frac{\tau_2}{\tau_1})} \approx \zeta_2-\zeta_1-2 \sqrt{\delta}$. Thus the maximum derivative with respect to the Schwarzschild parameter $r_s$ is $|\alpha| (\frac{k c}{n'}+|\alpha|^2 \chi) \frac{\delta L}{r_s c} (1+\frac{\tau_2}{\tau_1})$.

Putting all this together we are able to estimate the error bound of the Schwarzschild radius $r_s$. The variance of the estimator is:
\begin{equation}
\begin{split}
\frac{\braket{\Delta r_s^2}}{r_s^2}&= \frac{\braket{\Delta X^2}}{r_s^2(\frac{d\braket{X}}{d\tau} \frac{d\tau}{d r_s})^2}\\
&= \frac{\braket{\Delta X^2}}{r_s^2 |\alpha|^2 (\frac{k c}{n'}+|\alpha|^2 \chi)^2 (\frac{L}{c} \frac{\delta}{r_s})^2 (1+\frac{\tau_2}{\tau_1})^2}\\ 
&\approx\frac{1}{N(\frac{k c}{n'}+N \chi)^2 (\frac{L}{c})^2 (\frac{r_s h}{r_A r_B})^2 (1-\frac{r_s h}{2 r_A r_B})^2}
\end{split}
\end{equation}
Where $N=|\alpha|^2$ is the average number of coherent photons injected into the interferometer. Thus the relative error of the Schwarzschild radius $r_s$ of $M$ number of measurements is:
\begin{equation}
\begin{split}
\frac{\braket{\Delta r_s}}{r_s} = \frac{r_A r_B c}{L h r_s (1-\frac{r_s h}{2 r_A r_B}) \sqrt{M N(\frac{\omega}{n'}+N \chi)^2}}.
\end{split}
\end{equation}
This can be compared to the Fisher information bound obtained from Eq. \ref{CRbound} where the lower bound is exact.
\begin{equation}
\frac{\braket{\Delta r_s}_{opt}}{r_s} \ge \frac{r_A r_B c}{L h r_s \sqrt{ M N_a ((\omega+2(N_a+1)\chi)^2+2 N_a \chi^2)} }
\label{FB}
\end{equation}
Although the non-linear interferometer does not saturate the Fisher bound it does have the same photon number scaling for large intensities: $1/N^{3/2}$, which is beyond the usual Heisenberg limit.

\begin{figure}
	\includegraphics[width=0.9 \linewidth]{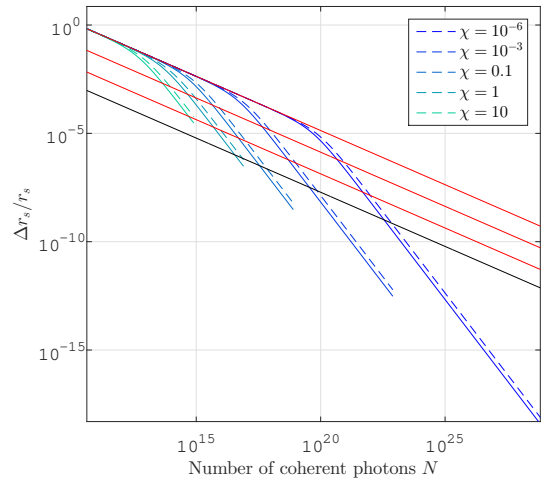}
	\caption{Error bound of the Schwarzschild radius plotted against number of coherent photons for various nonlinearity couplings of interferometer arm size $L=1$ cm and height $h=10$ m. The solid lines represent the exact lower bound for the best possible measurement. The dashed lines are the quadrature measurement error bounds. These lines terminate before the condition $|\alpha| \chi \tau <<1$ is violated. The solid black line is the case where squeezing of all photons is used to enhance the sensitivity, however small amounts of loss ($\epsilon=1-10^{-6}$) means the scaling is still at the SQL. From top to bottom, the red solid lines represent the SQL limit for a linear interferometer of heights $h=10$ m, $10^2$ m and $10^3$ m. (Other parameters: Number of measurements $M=10^{10}$, the central frequency $\omega=100$ THz and the radius $r_A=6.37 \times 10^6$ m (Earth's radius))}
	\label{rskerr}
\end{figure}

{\em Beyond-conventional-Heisenberg advantage for measuring space-time curvature}.-  We now wish to know at which point the scaling beyond the conventional Heisenberg limit becomes apparent. 
In Fig. \ref{rskerr}, we plot the optimized error bound of the Schwarzschild radius against the number of coherent photons for various non-linear couplings $\chi$. We have optimized this error with respect to the quadrature measurement angle. We have fixed the interferometer arm lengths to $L=1$ cm to ensure the condition $|\alpha|\chi \tau<<1$ for all values of $|\alpha| \chi$ in Fig. \ref{rskerr}. 

Furthermore, the height $h=10$ m with light at a central frequency of $\omega=100$ THz and $M=10$ GHz of measurements which are reasonable repetition rates \cite{thomas}. The $\propto \frac{1}{N^{3/2}}$ scaling becomes apparent for increasing number of photons $N$. As expected, for stronger coupling $\chi$, the scaling occurs for less number of photons. The quadrature measurement (dashed line) follows but never reaches the ultimate precision bound (Eq.\ref{FB}) represented by the solid line. We also plot the SNL for interferometer heights $h=10$ m, $10^2$ m and $10^3$ m represented by the red solid lines. For a pulse with $10^{18}$ photons, we'd only need $\chi=0.1$ for a precision of $10^{-8}$ which is a 4 order of magnitude improvement over the SQL scaling. State-of-the-art laser-cooled atom interferometry can measure gravity with a resolution of $2 \times 10^{-8}$ for a $1.3 s$ measurement \cite{peters}. However, this is limited to the SQL scaling. Future atom interferometers may be able to exploit entanglement resources to approach Heisenberg scaling and improve up to an order of $10^3$, as well as using a much longer measurement time \cite{savas}. Nonetheless, our optical scheme has the potential to outperform current state-of-the-art gravity measuring devices.  

By adding the Kerr non-linearities we reduce the area of the interferometer needed for a particular precision significantly. More generally, in terms of the unitless parameter $\tilde{y}=\frac{N \chi n'}{\omega}$ we find that the effect of the non-linearity becomes significant when $\tilde{y} \approx 1$, and dominates the scaling when $\tilde{y} \approx 100$. However, we have previously assumed the condition $\chi \tau \alpha=\tilde{y} \omega \frac{L}{c \sqrt{N}} << 1 \approx 0.01$. Therefore, for $N=10^{15}$, we have to limit the size of the nonlinearity to $L=\frac{0.01 c \sqrt{N}}{100 \omega}\approx 0.01$ m. Comparing the $h=10$ m non-linear noise limit and SQL, we see two or more orders of magnitude improvement equivalent to having a larger linear interferometer $h=10^3$ m. Thus by introducing the nonlinearity, we can downsize the interferometer size while keeping the precision the same. We note that the anti-squeezing noise for an error in the phase $\beta$ of $\Delta \beta=10^{-3}$ radians only changes $\braket{\Delta X^2}$ by $1$ dB (see Appendix C) and thus $\Delta r_s /r_s$ only increases an order of magnitude. Our scheme allows us to measure standard error in the phase of $10^{-10}$ radians in a single shot measurement, thus the added noise is negligible and doesn't affect $\Delta r_s/r_s$. 


{\em The effect of loss} - Whilst loss has a highly detrimental effect on the resolution improvements achieved via squeezing, it has a much smaller effect on the non-linear interferometer. We can model loss introduced due to non-unit detection efficiency via a beamsplitter of transmission $\epsilon_a$ after the non-linearities, and insertion losses on the probe via a beamsplitter of transmission $\epsilon_b$ before the non-linearities. These effects are straightforward to incorporate in the model (see Appendix E) giving the revised error bound:
%
%
\begin{equation}
\begin{split}
\frac{\braket{\Delta r_s}}{r_s} = \frac{r_A r_B c}{L h r_s (1-\frac{r_s h}{2 r_A r_B}) \sqrt{\epsilon_a \epsilon_b N(\frac{\omega}{n'}+ \epsilon_b N \chi)^2}}
\end{split}
\end{equation}
%
The loss reduces the effective size of the coherent amplitude but does not change the beyond-conventional-Heisenberg scaling.
In contrast, a squeezed coherent state will rapidly lose its non-classical properties through a lossy quantum channel. In Fig. \ref{rskerr} we have plotted for comparison the performance of an equivalent linear interferometer with squeezed light injected \cite{Gao}. As shown, the presence of a very small amount of loss keeps the scaling at the SQL whilst having virtually no effect on the non-linear interferometer.

{\em Experimental feasibility}.- Surpassing the conventional Heisenberg limit for the parameter $\chi$, rather than $\tau$ was recently demonstrated experimentally \cite{walmsley}. The energy scaling could be seen in a regime of low photon numbers by canceling the linear phase. Unlike our approach, quantum fluctuations were not considered and a strict condition of $\chi \tau N<<1$ was imposed, limiting the photon number to $N<10^8$. In our proposal, the values of the non-linearity $\chi$ and number of photons $N$ at which we get a significant improvement in the precision of $r_s$ are more challenging but may become available in the future. We note that the Kerr non-linearity constant depends on the pulse duration and the finite time of interaction of the single mode \cite{shapiro}. Our definition of $\chi$ describes an effective nonlinearity that is determined from classical theory (see Appendix D). For femto-second pulses in glass fibre the non-linearity is $\chi \approx 10^{-6}$ which would require over $10^{20}$ photons per pulse to see the enhancement. In Ref. \cite{nat}, $30$ femto-second pulses at $\omega=100$ THz frequency with $P=440$ GW peak power were produced, corresponding to $N=10^{18}$ photons per pulse, too low to observe the non-linear phase difference in glass fibre. However, in Ref \cite{matsuba}, pico-second pulses in photonic crystal fibres were shown to exhibit a much larger nonlinearity of $\chi \approx 6$ which implies from our results that over $N=10^{15}$ photons are needed. A further requirement is to ensure that the nonlinear material can withstand intense pulses without optical damage, Kerr saturation or plasma cladding \cite{bree, bastian, boyd}. 

{\em Conclusion}. We have studied the problem of estimating the phase shift due to the general relativistic time dilation in the interference of photons. We have identified that a non-linear interferometer with Kerr non-linearities $\chi$ in both arms couples to the space-time via a non-linear phase difference $\phi_{NL}$. The quantum error bound of the Schwarzschild radius was found to scale beyond the Heisenberg limit for a coherent probe state input. In principle, non-linear interactions of order $q \ge 2$ would scale $\propto \frac{1}{N^{q-\frac{1}{2}}}$. We analysed a sub-optimal quadrature measurement that nevertheless shows the same scaling. We found that our non-linear interferometer is more practical against loss compared to using squeezed coherent states. Finally, we believe that we are within reach of future experiments.      

\begin{acknowledgments}
{\em Acknowledgements}. This work was supported in part by the Australian Research Council Centre of Excellence for Quantum Computation and Communication Technology (Project No. CE110001027) and financial support by an Australian Government Research Training Program Scholarship.
\end{acknowledgments}

\appendix

\section{Calculation of coherent state overlap in equation 3}
 We consider the coherent state undergoing the non-linear evolution $U_{NL}=e^{-i \chi \tau (a^\dagger a)^2}$. To determine the fidelity $\mathcal{F}=|\braket{\alpha(\tau)|\alpha(\tau+d \tau)}|^2$ for a small change in the measured parameter $\tau$, we first determine the overlap:
\begin{equation}
\begin{split}
&\braket{\alpha_{NL}(\tau+d \tau)| \alpha_{NL}(\tau)}
=e^{-|\alpha|^2}  \sum_{n=0}^{\infty} \frac{(|\alpha|^2 e^{-i \chi d\tau (n+1)+i k c d \tau})^n}{n!} \\
&\approx e^{-|\alpha|^2} \sum_{n=0}^{\infty} \frac{|\alpha|^{2n} e^{i n k c d \tau}}{n!} (1-i \chi d\tau (n+1)n\\
&- \frac{\chi^2 d \tau^2 (n+1)^2 n^2}{2}) \\
&=e^{-|\alpha|^2(1-e^{i k c d \tau})} (1-i \alpha^2 e^{i \omega d\tau} (2+\alpha^2 e^{i \omega d\tau}) \chi d \tau \\
&-\frac{|\alpha|^2 e^{i \omega d\tau}}{2} (4 +14 e^{i \omega d\tau} |\alpha|^2+8 e^{2 i \omega d\tau} |\alpha|^4  \\
&+e^{3 i \omega d\tau} |\alpha|^6) \chi^2 d \tau^2)
\end{split}
\end{equation}

Expanding and only retaining terms up to second order in $d \tau$ gives Eq. 3 in the main text.

\section{Approximate Quantum Fisher Information for $q$ order non-linearity}
We want to determine the Cramer-Rao bound for $q$ order non-linear interaction with Hamiltonian $H=\chi (a^\dagger a)^q$.
We can approximate the unitary evolution using $a \approx |\alpha| + \delta a$ for very large coherent amplitude. Thus, the evolved coherent state becomes $e^{i\chi \tau (a^\dagger a)^q} \ket{\alpha} \approx e^{i \chi \tau |\alpha|^{2q}(1+\frac{q \delta a^\dagger}{|\alpha|})(1+\frac{q \delta a}{|\alpha|})} \ket{\alpha} \approx e^{i \chi \tau |\alpha|^{2q}} e^{i \chi \tau q |\alpha|^{2q-1}(\delta a^\dagger + \delta a)} e^{|\alpha|(\delta a^\dagger-\delta a)} \ket{0}\approx e^{i \chi \tau |\alpha|^{2q}} \ket{ \alpha (1+iq\chi \tau |\alpha|^{2(q-1)})}$. In general, for $q\ge 2$, 
\begin{equation}
\begin{split}
&\braket{\alpha_{NL}(\tau+d \tau)| \alpha_{NL}(\tau)}\\
&=e^{\frac{|\alpha|^2}{2} |(1-iq\chi (\tau+d\tau) |\alpha|^{2(q-1)})e^{i k c (\tau+d \tau)}-(1-iq\chi d \tau |\alpha|^{2(q-1)})e^{i k c \tau}|^2} \\
&\approx e^{-\frac{|\alpha|^2}{2} (q \chi |\alpha|^{2(q-1)} d\tau+k c d \tau)^2} \\
\end{split}
\end{equation}

Therefore, the fidelity is 
\begin{equation}
\mathcal{F}=1-N (q \chi N^{q-1}+k c)^2 d\tau^2
\end{equation}

And the Quantum Fisher information is:

\begin{equation}
\begin{split}
H(\tau)=4 N (q \chi N^{q-1}+k c)^2
\end{split}
\label{qfi0}
\end{equation}
\begin{figure}
	\hspace*{-0.1cm}
	\includegraphics[width=0.6 \linewidth]{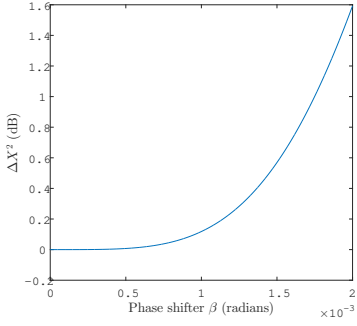}
	\caption{Quadrature noise at the chosen measurement angle $\theta=\frac{\pi}{2}-\zeta_2$. For approximately $\Delta \beta=\beta_a-\beta=1.5 \times 10^{-3}$ corresponding to a large systematic phase error, we only have an increase of $1$ dB of noise.}
	\label{quad}
\end{figure}
\section{Quadrature noise}
We consider the effect of how a systematic error in the choice of the phase $\beta$ can change the amount of noise. For the parameters $\chi=0.1$ and $N=|\alpha|^2=10^{17}$, we choose $\theta+\zeta_2= \frac{\pi}{2}$ and $\beta$ is the independent variable. As it turns out, for a small off-set from the optimum point of $10^{-3}$ radians in this $\beta$ phase, less than 1 dB of noise is added (see Fig. \ref{quad} graph). This doesn't seem to be a major issue since we predict a $\delta r_s/r_s = 10^{-3}$ and thus we can detect an absolute change of $10^{-10}$ radians in the phase for a single shot measurement. Therefore, a large systematic error doesn't add significant noise to destroy the beyond-conventional-Heisenberg scaling. 

\section{Experimental feasibility}
\begin{figure}
	\hspace*{-0.1cm}
	\includegraphics[width=0.6 \linewidth]{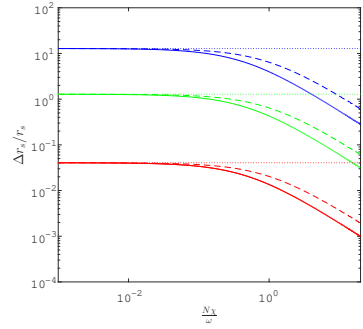}
	\caption{Error bound of the Schwarzschild radius plotted against the unitless quantity $\frac{N \chi}{\omega}$ for various $N$ and fixed length $L=1000$ m and $h=1$ m. From top to bottom, each colour represents $N=10^{15}$, $N=10^{17}$ and $N=10^{20}$. The solid lines represent the theoretical quantum error bound. The dashed represent the quadrature measurement. The dotted lines represent the shot noise limit for a linear interferometer. (Other parameters are $M=10^{10}$)}
	\label{opt}
\end{figure}

In Fig. \ref{opt}, we present the relative Schwarzschild error bound plotted against the unitless parameter $\tilde{y}=\frac{N \chi n'}{\omega}$. Thus, we can rewrite the error bounds as:

\begin{equation}
\begin{split}
\frac{\braket{\Delta r_s}}{r_s} = \frac{r_A r_B c n'}{L h r_s \omega (1-\frac{r_s h}{2 r_A r_B}) \sqrt{M N(1+\tilde{y})^2}}
\end{split}
\end{equation}

And

\begin{equation}
\frac{\braket{\Delta r_s}_{opt}}{r_s} \ge \frac{r_A r_B c n'}{2 L h r_s \omega \sqrt{M N_a ((1+2 \tilde{y}+2\chi)^2+2 \tilde{y} \chi)} }
\end{equation}

Where $M$ is the number of single shot measurements.
From these expressions, we expect that the turning point at which the non-linearity becomes significant is approximately when $\tilde{y} \approx 10$. As seen in Fig. \ref{opt}, for a fixed number of photons $N$ and central frequency $\omega$, there is approximately an order of magnitude improvement over a SNL linear interferometer. A conservative estimate of $\chi$ for $N=10^{15}$, $10^{17}$, $10^{20}$ respectively is $\chi= \frac{\tilde{y} \omega}{N}=1$, $10^{-2}$ and $10^{-5}$. Let's consider the case of $\chi=10^{-5}$ for which the number of photons per $\Delta t=30$ $fs$ pulse duration is $N=10^{20}$ with $M=10^{10}$ number of measurements would correspond to a peak power of $P=\frac{N \hbar \omega}{\Delta t}  \approx 4 \times 10^{13}$ W=$40$ TW (Average power $\tilde{P}=10$ GW). On the other hand, for a stronger linearity of $\chi=1$, the peak power required to see the enhancement with $N=10^{15}$ photons per pulse would reduce to $P=400$ MW and an average power of $\tilde{P}=100$ kW. We note similarities in these values with Ref. \cite{S_LUIS}. 

The definition of the nonlinearity constant $\chi'$ in Ref. \cite{S_LUIS} is slightly different from our definition. Namely, $\chi'$ represents the phase shift per unit photon. It is defined as:

\begin{equation}
\chi'=\frac{\tilde{n}}{n_0} \frac{\hbar \omega}{A \Delta t}
\end{equation}
Where $\tilde{n}$ is the second order refractive index from the expansion $n=n_0+\tilde{n} I$, $A$ is the area of the pulse, and $\Delta t$ is its duration. Thus, the nonlinear phase shift per photon can be increased by reducing the area and the pulse duration. It follows that the phase shift is given by $\phi'_{NL}=\frac{n_0 \omega L}{c} \frac{\chi'}{2} N$. Comparing with our phase shift $\phi_{NL}=\frac{L}{c} \chi N$, the relation between our non-linear coefficient and that in Ref \cite{S_LUIS} is $\chi= \frac{n_0}{2}  \omega \chi'$.

The values of the nonlinearities quoted in the main text are based on converting the given formula of the phase $\phi_{NL}= |\alpha|^2 \chi \tau$ from the values given. For example, a nonlinear phase shift of $10^{-8}-10^{-7}$ with the given fibre length of $L=4.5$ m in Ref. \cite{matsuba} for a single photon correponds to $\chi=1$ to $\chi=6$. The same calculation was done for the optical fibre. 

\section{Including loss}

{\em The effect of loss on the non-linear interferometer} - Whilst loss has a highly detrimental effect on the resolution improvements achieved via squeezing, it has a much smaller effect on the non-linear interferometer. We can model loss introduced due to non-unit detection efficiency via a beamsplitter of transmission $\epsilon_a$ after the non-linearities, and insertion losses on the probe via a beamsplitter of transmission $\epsilon_b$ before the non-linearities. These effects are straightforward to incorporate in the model giving the revised error bound:

Loss after the non-linearity leads to $e^{-i a^\dagger a \chi \tau} a \rightarrow e^{-i a^\dagger a \chi \tau} \sqrt{\epsilon_a} a+\sqrt{1-\epsilon_a} d $ and after the beamsplitter becomes:
\begin{equation}
\begin{split}
X_b&=b(\tau) e^{i\theta}+b^\dagger(\tau)e^{-i\theta}\\
&=\sqrt{\epsilon_a} |\alpha| \cos{(\theta+\zeta_2}) - \sqrt{\epsilon_a} |\alpha| \cos{(\theta+\zeta_1+\beta)} \\
&-\chi |\alpha|^2 (\tau_2 \sin{(\theta+\zeta_2)}-\tau_1 \sin{(\theta+\zeta_1+\beta)}) \sqrt{\epsilon_a} \delta X_a \\
&+\frac{\sqrt{\epsilon}}{2}(\delta X_{a(\theta+\zeta_2)}-\delta X_{a (\theta+\zeta_1+\beta}))\\
&+\frac{\sqrt{1-\epsilon}}{\sqrt{2}}(\delta X_{d(\theta+\zeta_2)}-\delta X_{d'(\theta+\zeta_1+\beta)})\\
& +\frac{\sqrt{\epsilon}}{2}(X_{v(\theta+\zeta_2)}+X_{v (\theta+\zeta_1+\beta)})
\label{xe}
\end{split}
\end{equation}
And the variance is:
\begin{equation}
\begin{split}
&\braket{\Delta X_b^2}=\epsilon_a \chi^2 |\alpha|^4 (\tau_2 \sin{(\theta+\zeta_2)}-\tau_1 \sin{(\theta+\zeta_1+\beta)})^2\\
&-\epsilon_a \chi |\alpha|^2 (\tau_2 \sin{(\theta+\zeta_2)}-\tau_1 \sin{(\theta+\zeta_1+\beta)}) \\
& \times (\cos{(\theta+\zeta_2)}-\cos{(\theta+\zeta_1+\beta)})+1\\
\end{split}
\end{equation} 
For the optimal angle, the variance reduces also to shot noise $\braket{\Delta X^2}=1$. Loss before the non-linearities simply reduces the input photon number by the factor $\epsilon_b$. Therefore, the error bound for the combined case of having loss before and after the non-linearities is:
\begin{equation}
\begin{split}
\frac{\braket{\Delta r_s}}{r_s} = \frac{r_A r_B c}{L h r_s (1-\frac{r_s h}{2 r_A r_B}) \sqrt{\epsilon_a \epsilon_b N(\frac{\omega}{n'}+ \epsilon_b N \chi)^2}}
\end{split}
\end{equation}
%
The loss reduces the effective size of the coherent amplitude but does not change the super-Heisenberg scaling.
In contrast, a squeezed coherent state will lose its non-classical properties through a lossy quantum channel. In Fig.2 of the main text we have plotted for comparison the performance of an equivalent linear interferometer with squeezed light injected \cite{S_Gao}. As shown, the presence of a very small amount of loss destroys the advantage of the squeezing whilst having virtually no effect on the non-linear interferometer.
The ultimate limit for a lossy interferometer with squeezed coherent probe states is  \cite{S_Gao}:
\begin{equation}
\begin{split}
\frac{\braket{\Delta{r_s}}}{r_s} & \ge \frac{r_A r_B c n'}{2 L h r_s \omega \sqrt{\frac{\epsilon N_c}{1-\epsilon+\epsilon e^{-2 r}}+\epsilon N_s}}
\end{split}
\end{equation}
Where $N_c$ and $N_s$ is the number of coherent and squeezed photons, respectively. We assume the squeezing parameter $r$ is positive and very large. Consequently, for significant loss $\epsilon << 1$, the Heisenberg scaling of $\propto \frac{1}{N}$ is lost for the optimal number of squeezed photons $N_s=N_c$ and reduces to the SNL. Loss on the order of $\epsilon \approx 1-\frac{1}{N_{\chi}}$ where $N_{\chi}$ is the turning point of the scaling for the respective value of the non-linearity is enough to destroy the Heisenberg scaling as seen in Fig. [2] of the main text. On the other hand, our non-linear interferometer setup requires only a $\frac{1}{\epsilon}$ increase in the input number of coherent photons to compensate for the loss.


\begin{thebibliography}{99}

\bibitem{GIO11} {\it Advances in quantum metrology}, V. Giovanetti, S. Lloyd, and L. Maccone, Nature Photon. {\bf 5}, 222 (2011).

\bibitem{ABB16} {\it Observation of Gravitational Waves from a Binary Black Hole Merger},
B. P. Abbott et al., Phys. Rev. Lett. {\bf 116}, 061102 (2016).

\bibitem{CAV81} {\it Quantum-mechanical noise in an interferometer},
C. M. Caves,
Phys. Rev. D {\bf 23}, 1693 (1981).

\bibitem{BOI07} {\it Breaking the Heisenberg limit with inefficient detectors},
J. Beltr\'an and A. Luis,
Phys. Rev. A {\bf 72}, 045801 (2005) 
\bibitem{BOI} {\it Generalised limits for single-parameter quantum estimation},
S. Boixo, S. T. Flammia, C. M. Caves, Phys. Rev. Lett., {\bf 98},  090401 (2007).


\bibitem{HAL12} {\it Does Nonlinear Metrology Offer Improved Resolution? Answers from Quantum
	Information Theory},
M. J. W. Hall and H. M. Wiseman, Phys. Rev. X {\bf 2}, 041006 (2012).

\bibitem{zwierz} {\it General optimality of the Heisenberg limit for quantum metrology}, Marcin Zwierz, Carlos A. Perez-Delgado, Pieter Kok, Phys. Rev. Lett. {\bf 105}, 180402 (2010).


\bibitem{NAP11} {\it Interaction-based quantum metrology showing scaling beyond the Heisenberg limit},
M.Napolitano, et al, Nature, {\bf 471},  486 (2011).

\bibitem{JOO12} {\it Quantum metrology for non-linear phase shifts with entangled coherent states},
Jaewoo Joo, Kimin Park, Hyunseok Jeong, William J. Munro, Kae Nemoto, Timothy P. Spiller,
Phys. Rev. A {\bf 86}, 043828 (2012).


\bibitem{LUIS} {\it Nonlinear Michelson interferometer for improved quantum metrology}, A. Luis, A. Rivas, Phys. Rev. A {\bf 92}, 022104 (2015).

\bibitem{zych} {\it General relativistic effects in quantum interference of photons}, Magdalena Zych, Fabio Costa, Igor Pikovski, Timothy C. Ralph and Caslav Brukner, Class. Quantum Grav. {\bf 29} 224010 (2012).

\bibitem{peters} {\it High-precision gravity measurements using atom interferometry}, A. Peters, K. Y. Chung, S. Chu, Metrologia {\bf 38}, 25 (2001).

\bibitem{savas} {\it Testing General Relativity with Atom Interferometry}, S. Dimopoulos, P. W. Graham, J. M. Hogan, M. A. Kasevich, Phys. Rev. Lett. {\bf 98}, 111102 (2007).

\bibitem{szigeti} {\it Squeezed-light-enhanced atom interferometry below the standard quantum limit}, S. S. Szigeti, B. Tonekaboni, W. Y. S. Lau, S. N. Hood, and S. A. Haine, Phys. Rev. A {\bf 90}, 063630 (2014).

\bibitem{esteve} {\it Squeezing and entanglement in a Bose Einstein condensate}, J. Esteve, C. Gross, A. Weller, S. Giovanazzi, M. K. Oberthaler, Nature {\bf 455}, 1216-1219 (2008)

\bibitem{gross} {\it Nonlinear atom interferometer surpasses classical precision limit}, C. Gross, T. Zibold, E. Nicklas, J. Estève, M. K. Oberthaler, Nature {\bf 464}, 1165-1169 (2010)

\bibitem{qft} {\it Quantum Field Theory}, M. Srednicki, Cambridge University Press pg. 10 (2007).

\bibitem{milburn} {\it Quantum Optics}, D. F. Walls, Gerard J. Milburn, 2nd Edition Springer-Verlag Berlin Heidelberg (2008).

\bibitem{cram} {\it Mathematical methods of statistics}, H. Cramer, Princeton University Press, (1946).

\bibitem{mras} {\it Optimal Quantum Estimation of Loss in Bosonic Channels}, A. Monras and M. G. A. Paris, Phys. Rev. Lett. {\bf 98},
160401 (2007).

\bibitem{introqfi} {\it Quantum Information: An Introduction}, M. Hayashi, Berlin: Springer (2006)

\bibitem{qfidef} {\it Heisenberg scaling in Gaussian quantum metrology}, N. Friis, M. Skotiniotis, I. Fuentes, and W. Dur, Phys. Rev. A {\bf 92}, 022106 (2015)

\bibitem{safranek} {\it Quantum parameter estimation using multi-mode Gaussian states} D. Safranek, A. R. Lee, and I. Fuentes, New J. Phys. {\bf 17}, 073016
(2015)

\bibitem{BRA94} {\it Statistical distance and the geometry of quantum states},
S. L. Braunstein and C. M. Caves,
Phys. Rev. Lett. {\bf 72}, 3439 (1994).

\bibitem{strobel} {\it Quantum Metrology. Fisher Information and Entanglement of Non-Gaussian Spin States}, H. Strobel, W. Muessel, D. Linnemann, T. Zibold, D. B.
Hume, L. Pezz, A. Smerzi, and M. K. Oberthaler, Science
{\bf 345}, 424 (2014) 


\bibitem{dodonov} {\it Theory of Nonclassical States of Light, Chapter 6: Nonclassical states of light propagating in Kerr media}, pg. 285, V. V. Dodonov, V. I. Man'ko, R. Tanas, Taylor \& Francis (2003).

\bibitem{shapiro} {\it Quantum propagation in a Kerr medium: lossless, dispersionless fiber}, L. G. Joneckis and J. H. Shapiro, J. Opt. Soc. Am. B {\bf 10}, 1102 (1993).

\bibitem{kit} {\it Number-phase minimum-uncertainty state with reduced number uncertainty in a Kerr nonlinear interferometer}, M. Kitagawa and Y. Yamamoto, Phys. Rev. A {\bf 34}, 3974 (1986).

\bibitem{thomas} {\it 10-GHz, 1.3-ps erbium fiber laser employing soliton pulse shortening}, T. F. Carruthers, I. N. Duling, Optics Letters {\bf 21}, 23:1927-1929 (1996).





\bibitem{Gao} {\it Quantum optical metrology in the lossy SU(2) and SU(1,1) interferometers}, Y. Gao, Phys. Rev. A {\bf 94}, 023834 (2016).

\bibitem{walmsley} {\it Surpassing the conventional Heisenberg limit using classical resources},  X. Jin, M. Lebrat, L. Zhang, K. Lee, T. Bartley, M. Barbieri, J. Nunn, A. Datta, and I. A. Walmsley, CLEO:
QELS Fundamental Science 2013 paper QF2B.2, San Jose, CA, OSA Technical
Digest (online) (OSA, Washington, D.C., 2013).

\bibitem{nat} {\it Multi-millijoule few-cycle mid-infrared pulses through nonlinear self-compression in bulk}, V. Shumakova, P. Malevich, S. Alisauskas, A. Voronin, A. M. Zheltikov, D. Faccio, D. Kartashov, A. Baltuška, and A. Pugžlys, Nat. Commun. {\bf 7}:12877 (2016).


\bibitem{matsuba} {\it Observation of optical-fibre Kerr nonlinearity at the single-photon level}, N. Matsuda, R. Shimizu, Y. Mitsumori, H. Kosaka, K. Edamatsu, Nat. Photon. {\bf 3}, 95 (2009).

\bibitem{bree} {\it Saturation of the all-optical Kerr Effect}, C. Bree, A. Demircan, G. Steinmeyer, Phys. Rev. Lett. {\bf 106}, 183902 (2011).

\bibitem{bastian} {\it Saturation of the all-optical Kerr effect in solids}, B. Borchers, C. Bree, S. Birkholz, A. Demircan, G. Steinmeyer, Optics Letters {\bf 37} 9 (2012).

\bibitem{boyd} {\it Nonlinear Optics (Third Edition)}, Chap 4.7, R. W. Boyd, Elsevier (2008).

\end{thebibliography}

\begin{thebibliography}{99}
	
\bibitem{S_LUIS} {\it Nonlinear Michelson interferometer for improved quantum metrology}, A. Luis, A. Rivas, Phys. Rev. A {\bf 92}, 022104 (2015)	

\bibitem{matsuba} {\it Observation of optical-fibre Kerr nonlinearity at the single-photon level}, N. Matsuda, R. Shimizu, Y. Mitsumori, H. Kosaka, K. Edamatsu, Nat. Photon. {\bf 3}, 95 (2009)


	\bibitem{S_Gao} {\it Quantum optical metrology in the lossy SU(2) and SU(1,1) interferometers}, Y. Gao, Phys. Rev. A {\bf 94}, 023834 (2016)
	
	
\end{thebibliography}
\end{document}